\documentclass[conference]{IEEEtran}
\IEEEoverridecommandlockouts
\usepackage{cite}
\usepackage{amsmath,amssymb,amsfonts}
\usepackage{algorithmic}
\usepackage{graphicx}
\usepackage{textcomp}
\usepackage{xcolor}
\usepackage{url}
\usepackage{booktabs}
\usepackage{multirow}
\usepackage{array}
\usepackage{threeparttable}
\usepackage[hidelinks]{hyperref}
\hypersetup{
  pdftitle={CUJBench: Benchmarking LLM Agents for Browser-to-Backend Failure Diagnosis},
  pdfauthor={Haoming Meng},
  pdfkeywords={LLM agents, multi-modal evaluation, critical user journeys, failure diagnosis, AIOps, software reliability, benchmark, root cause analysis}
}
\def\BibTeX{{\rm B\kern-.05em{\sc i\kern-.025em b}\kern-.08em
    T\kern-.1667em\lower.7ex\hbox{E}\kern-.125emX}}
\begin{document}

\title{CUJBench: Benchmarking LLM Agents for Browser-to-Backend Failure Diagnosis}

\author{\IEEEauthorblockN{Haoming Meng}
\IEEEauthorblockA{\textit{Snowflake}\\
Bellevue, WA, USA\\
haoming.meng@snowflake.com}
}

\maketitle

\begin{abstract}
Automated failure diagnosis requires correlating browser-visible symptoms with
backend observability signals, yet existing benchmarks do not evaluate this
cross-modal reasoning task. Constructing one is non-trivial: multi-modal
failure scenarios are costly to annotate, and live-environment capture
introduces stochasticity that makes cross-run agent comparison unreliable.
We present \textbf{CUJBench}, to our knowledge, the first benchmark to combine
browser-visible failure evidence with backend observability in a diagnostic
framing. CUJBench addresses annotation cost through an LLM-assisted generation
pipeline with a multi-agent review loop and a three-layer annotation scheme,
producing 87 labeled scenarios across five fault families, and ensures
reproducibility by packaging each failure as a deterministic multi-modal
snapshot with a fixed tool interface. Evaluating six LLMs under
retrieval, browser-only, and full-toolset baselines, the benchmark yields an
overall accuracy of 21.6\% with a ceiling of 52.5\%, well below saturation.
Contrary to expectation, browser-only agents outperform full-toolset agents in
aggregate, with expanded evidence access inducing unfocused exploration rather
than improved synthesis. Trajectory analysis identifies cross-modal synthesis
as the primary bottleneck: agents retrieve the decisive evidence but fail to
attribute it correctly---a failure that persists across all six models and that
model scale and richer tool access alone cannot resolve.
\end{abstract}


\begin{center}
\begin{minipage}{0.96\columnwidth}
\scriptsize
\copyright~2026 IEEE. Personal use of this material is permitted. Permission
from IEEE must be obtained for all other uses, in any current or future media,
including reprinting/republishing this material for advertising or promotional
purposes, creating new collective works, for resale or redistribution to
servers or lists, or reuse of any copyrighted component of this work in other
works.
\end{minipage}
\end{center}

\section{Introduction}
\label{sec:intro}

Large language model (LLM) agents have emerged as a compelling approach
to automating cloud operations.
Recent frameworks such as AIOpsLab~\cite{b1}, ITBench~\cite{b2}, and
Cloud-OpsBench~\cite{b3} have pushed the frontier of agentic root cause
analysis (RCA), equipping agents with tool interfaces to query
telemetry, hypothesize faults, and trace causal chains through microservice
environments.
Yet the performance ceiling remains low: across five frontier models on the
OpenRCA benchmark, perfect RCA accuracy ranges from only 3.9\% to
12.5\%~\cite{b8}, with dominant failure modes that originate from agent
framework structure rather than from individual model limitations.
The community has responded with increasingly sophisticated evaluation
protocols, from outcome-based scoring to trajectory-alignment and
process-level analysis~\cite{b3,b12}, consistent with broader recent
surveys of LLM-agent evaluation that emphasize realistic, fine-grained,
and scalable assessment of interactive agents~\cite{b9,b14}.
This difficulty is consistent with empirical studies of production-cloud
incidents, which show that incident response remains costly across
detection, root-causing, and mitigation stages even in mature
large-scale services~\cite{b30}.
Despite this progress, the benchmark landscape has a structural blind spot:
every existing benchmark evaluates agents on backend observability alone.

\textbf{Gap~1: The browser-visible layer of failure diagnosis is absent from
every RCA and AIOps benchmark.}
Modern software reliability practice relies on end-to-end (e2e) functional
tests that continuously validate that key product flows---login, checkout,
search---work as expected from the user's perspective. These end-to-end test flows covering the most business-critical product interactions are known as Critical User Journeys (CUJs)~\cite{b46,b47}. When a CUJ fails, diagnosis naturally begins with user-facing evidence:
a screenshot capturing the broken page state, a network waterfall showing the
failing HTTP request, or a JavaScript console exception. Only after these browser-visible artifacts surface the symptom does the engineer reach into backend observability to locate the root cause. This cross-modal reasoning chain, from browser-visible symptoms to backend
signals, is absent from every existing benchmark.
AIOpsLab~\cite{b1}, ITBench~\cite{b2}, Cloud-OpsBench~\cite{b3},
OpenRCA~\cite{b24}, and RCAEval~\cite{b25} evaluate agents exclusively on
backend telemetry. Even the widest multimodal RCA dataset, AnoMod~\cite{b42}, extends only to API response bodies and code coverage reports, which are still backend-side signals. No benchmark evaluates whether LLM agents can diagnose failures from CUJ tests.

\textbf{Gap~2: Web and GUI agent benchmarks evaluate task completion on
functioning applications, not failure diagnosis.}
Benchmarks such as WebArena~\cite{b5}, VisualWebArena~\cite{b4}, and
OSWorld~\cite{b6}, along with more recent long-horizon suites such as
TheAgentCompany~\cite{b43}, OdysseyBench~\cite{b44}, and The Tool
Decathlon~\cite{b45}, use browser-visible or tool-mediated evidence but frame
every task as completion on a functioning application. The diagnostic inversion---\textit{``X just failed; why?''}---is absent from
this literature. CUJ failure diagnosis demands exactly this inversion, and it combines the browser-visible evidence of web agent tasks with the backend observability of RCA benchmarks in a single multi-modal diagnostic problem.

\textbf{Gap~3: No benchmark combines browser and
backend evidence with process-level evaluation of diagnostic reasoning.}
Cloud-OpsBench~\cite{b3} introduced deterministic snapshot evaluation and
trajectory-alignment scoring, but its evidence surface is backend-only.
Recent process-level analyses reveal that incomplete evidence exploration and
hallucinated data interpretation are the dominant failure modes even in
backend-only RCA~\cite{b8,b12}; a benchmark that adds browser-visible
evidence to the diagnostic surface must also evaluate whether agents correctly
gather and integrate evidence across both modalities, not merely whether they
identify the correct root-cause label.
Building such a benchmark poses non-trivial challenges: annotating failure
scenarios across heterogeneous artifact types is labor-intensive~\cite{b51},
and live-environment capture introduces stochasticity that makes cross-run
comparison unreliable~\cite{b1,b2,b18,b25}.
Whether frontier models such as Claude Sonnet~4.6~\cite{b15} and
Gemini~3.1~Pro~\cite{b16} can perform this cross-modal reasoning in practice
therefore remains untested.

To bridge these three gaps, we present \textbf{CUJBench}, to our knowledge
the first benchmark to combine browser-visible failure evidence with backend
observability signals in a diagnostic task framing.
CUJBench adopts the deterministic snapshot paradigm of
Cloud-OpsBench~\cite{b3}: each scenario is a reproducible, immutable snapshot
of a CUJ test failure, packaged with a fixed tool interface
backed by pre-computed frontend and backend evidence caches.
The benchmark comprises 87 scenarios spanning five fault
families across two complementary open-source full-stack e-commerce applications: OpenTelemetry
Demo~\cite{b17}, which contributes backend-observable and cross-modal failures
on a polyglot microservices stack, and Tractor Store~\cite{b50}, which contributes
browser-dominant and micro-frontend failures in which screenshots, HAR data,
and browser-side spans often carry the decisive diagnostic signal while backend
telemetry is sparse.

The benchmark's scope, methodology, and evaluation protocol collectively address each gap identified above; we report the following contributions:

\begin{enumerate}

  \item \textbf{CUJBench} (Gap~1 and Gap~2): the
  first diagnostic benchmark to combine browser-visible failure evidence with
  backend observability in a single task.

  \item \textbf{Multi-modal snapshot methodology} (Gap~1): a scenario generation harness combining controlled fault injection with LLM-assisted authoring verified by a multi-agent review loop, producing deterministic snapshots across backend-observable, cross-modal, and browser-dominant diagnostic regimes.

  \item \textbf{Layered ground-truth and process-evaluation protocol}
  (Gap~3): structured root-cause annotations with cited evidence identifiers,
  reference diagnostic trajectories, and a modality-aware complexity taxonomy
  enabling deterministic scoring on component/layer/type accuracy,
  evidence-citation recall, and trajectory coverage.

\end{enumerate}

The benchmark harness and scenario artifacts are publicly available at \url{https://github.com/haoming29/CUJBench}.

\section{CUJ Bench}

\begin{figure*}[t]
\centering
\includegraphics[width=\textwidth]{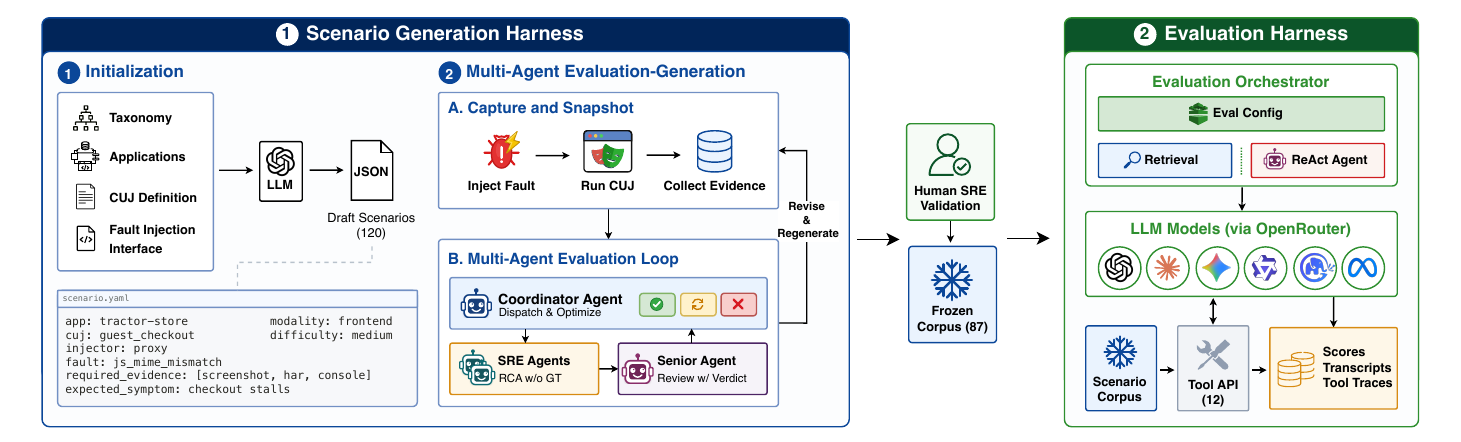}
\caption{Overview of CUJBench.}
\label{fig:cujbench-e2e}
\end{figure*}

CUJBench instantiates the diagnosis task in a reproducible benchmark harness
that couples instrumented open-source applications, scripted CUJ tests,
controlled fault injection, and deterministic snapshot packaging.
Rather than evaluating agents on live systems at inference time, the harness
captures each failure once under controlled conditions and exposes the resulting
evidence through a fixed, replayable tool interface---making every agent run
over the same scenario strictly comparable.
Table~\ref{tab:scenario-taxonomy} summarizes the corpus; Figure~\ref{fig:cujbench-e2e}
traces the end-to-end workflow from taxonomy-driven scenario construction and
multi-agent curation to frozen-snapshot evaluation.

\begin{table*}[t]
\caption{Scenario taxonomy of CUJBench corpus}
\label{tab:scenario-taxonomy}
\centering
\small
\renewcommand{\arraystretch}{1.12}
\setlength{\tabcolsep}{6pt}
\begin{tabular}{@{}>{\raggedright\arraybackslash}p{3.0cm}>{\raggedright\arraybackslash}p{2.1cm}>{\raggedright\arraybackslash}p{6.8cm}cccc@{}}
\toprule
\multirow{2}{*}{\textbf{Family}} & \multirow{2}{*}{\textbf{Apps}} & \multirow{2}{*}{\textbf{Representative Signal}} & \multicolumn{3}{c}{\textbf{Difficulty}} & \multirow{2}{*}{\textbf{N}} \\
\cmidrule(lr){4-6}
 & & & \textbf{E} & \textbf{M} & \textbf{H} & \\
\midrule
Baseline &
Both &
Healthy end-to-end CUJs with no injected failure &
2 & 0 & 0 & 2 \\
Browser proxy faults &
Both &
HAR, screenshots, DOM, console, and browser-visible timing or content anomalies &
5 & 48 & 3 & 56 \\
Backend flag faults &
OTel Demo &
Recent changes, traces, logs, metrics, and backend service symptoms &
0 & 4 & 0 & 4 \\
Compound faults &
OTel Demo &
Cross-modal evidence linking browser symptoms with backend change or telemetry signal &
0 & 0 & 18 & 18 \\
Frontend mutations &
Tractor Store &
Event-flow drift, listener failures, and browser state inconsistencies &
0 & 3 & 4 & 7 \\
\bottomrule
\end{tabular}
\end{table*}

\subsection{Problem Definition}

\label{sec:problem_def}

CUJBench formulates failed-CUJ diagnosis as a structured inference task over a
frozen multi-modal incident snapshot. Given a benchmark instance, the agent
must infer why a failed CUJ test broke from the evidence captured around that failure.

\begin{equation}
(s,\; y^{*}) \;\in\; \mathcal{S} \times \mathcal{Y}
\label{eq:problem_instance}
\end{equation}

Each benchmark instance consists of a scenario snapshot $s$ and a hidden
ground-truth annotation $y^{*}$. The scenario space $\mathcal{S}$ contains the
observable inputs available to the agent, while $\mathcal{Y}$ contains the
complete ground-truth annotation used for evaluation. A scenario snapshot packages the
evidence of a single CUJ failure:

\begin{equation}
s = \langle\, a,\; \mathcal{E} \,\rangle
\label{eq:scenario_snapshot}
\end{equation}

where $a$ is the incident alert and $\mathcal{E}$ is the pre-captured
multi-modal evidence. The evidence is partitioned by modality:

\begin{equation}
\mathcal{E} = \mathcal{E}_{f} \;\cup\; \mathcal{E}_{b} \;\cup\; \mathcal{E}_{c}
\label{eq:evidence}
\end{equation}

where $\mathcal{E}_{f}$ (frontend) captures browser session artifacts recorded
during the CUJ run (e.g., Playwright test report and HAR file);
$\mathcal{E}_{b}$ (backend) covers service-level observability such as
metrics, traces and logs; and $\mathcal{E}_{c}$ (context) provides
operational context such as recent changes and service topology. This three-way partition is the defining
structural feature of CUJBench: existing AIOps benchmarks operate primarily on
$\mathcal{E}_{b}$, while web-agent benchmarks operate primarily on
$\mathcal{E}_{f}$.

The agent does not receive the raw evidence $\mathcal{E}$ directly; instead it
accesses $s$ through a fixed tool interface $\mathcal{T}$, where each tool
returns a deterministic cached response from the scenario snapshot. The agent is modeled as a mapping from observable scenario inputs to
a structured RCA:

\begin{equation}
f : \mathcal{S} \;\longrightarrow\; \hat{\mathcal{Y}}, \qquad \hat{y} = f(s)
\label{eq:task_mapping}
\end{equation}

\begin{equation}
\hat{y} = (\hat{c},\; \hat{\ell},\; \hat{t},\; \hat{d},\; \hat{I})
\label{eq:prediction_space}
\end{equation}

The predicted diagnosis $\hat{y}$ contains the inferred faulty component
$\hat{c}$, fault layer $\hat{\ell}$, root-cause type $\hat{t}$,
natural-language reasoning $\hat{d}$, and cited evidence identifiers $\hat{I}$,
a finite set of artifact identifiers drawn from the captured evidence:

\begin{equation}
\hat{I} \;\subseteq\; \{\,\mathrm{id}(e) : e \in \mathcal{E}\,\}
\label{eq:evidence_ids}
\end{equation}

where $\mathrm{id}(e)$ is the unique identifier of artifact $e$
(e.g., {\small\texttt{har:req\_0030}}, {\small\texttt{log:frontend:0123}}), representing the
set of artifacts the agent judges sufficient to support its conclusion.

Note that $\hat{\mathcal{Y}} \subsetneq \mathcal{Y}$: the prediction space omits
$\tau^{*}$, since agents submit a diagnosis but not a trajectory. The agent's
tool-call sequence $\hat{\tau}$ is recorded as a side effect of
evaluation and compared against the human-annotated $\tau^{*}$ for process
metrics.

The ground truth label $y^{*}$ provides reference values for each field in
$\hat{y}$ and additionally includes $\tau^{*}$, the annotated reference
investigation trajectory:

\begin{equation}
y^{*} = (c^{*},\; \ell^{*},\; t^{*},\; d^{*},\; I^{*},\; \tau^{*})
\label{eq:ground_truth}
\end{equation}

\begin{equation}
\tau^{*} = \langle\,(\sigma_1, r_1),\;\ldots,\;(\sigma_n, r_n)\,\rangle, \quad \sigma_i \in \mathcal{T}
\label{eq:ref_trajectory}
\end{equation}

where each step $(\sigma_i, r_i)$ pairs a mandatory tool $\sigma_i$ with its
investigative rationale $r_i$. The mandatory tool set
$M = \{\sigma_i : (\sigma_i, r_i) \in \tau^{*}\}$ and the acceptable exploration
set $A \subseteq \mathcal{T}$ are derived from the annotation for process
evaluation. The ordering in $\tau^{*}$ is preserved in the annotation; current
evaluation metrics treat $M$ and $A$ as unordered sets---ordered trajectory
alignment is left as future work.

This formulation captures two properties that distinguish CUJBench from prior
benchmarks. First, the task is \textit{diagnostic} rather than
task-completion-oriented: the failure has already occurred and the goal is to
explain it. Second, the task is inherently \textit{multi-modal}: depending on
the scenario, the decisive signal may lie in $\mathcal{E}_{f}$,
$\mathcal{E}_{b}$, $\mathcal{E}_{c}$, or only in their combination.

\subsection{Design Decisions}

The gaps identified in Section~\ref{sec:intro} motivate three foundational
design decisions that directly shape the harness.

\textbf{Snapshot-based evaluation} (Gaps~1--3).
Benchmarks that require a live instrumented environment at inference time introduce
stochasticity through service startup variability, background noise, and dynamic
cluster state~\cite{b1,b2,b18,b25}, making cross-run comparison of agent behavior
unreliable. Cloud-OpsBench~\cite{b3} introduced scenario snapshots for backend RCA
but does not extend this to browser-side evidence. CUJBench generalizes the
snapshot discipline across the full evidence surface: each CUJ failure is captured
once under controlled conditions and packaged into a fixed artifact; evaluation
operates exclusively over the frozen package, eliminating live-cluster dependencies
and making every run over the same scenario strictly comparable. This discipline is
also a prerequisite for the process-level evaluation demanded by Gap~3: comparing
agent investigation trajectories is only meaningful when every run of the same
scenario sees the same alert, the same artifacts, and the same tool responses.

\textbf{Two complementary diagnostic regimes} (Gaps~1 and~2).
Existing diagnosis benchmarks occupy two non-overlapping evidence spaces.
Microservice RCA work---including methods that explicitly target multi-modal
observability---operates within the backend stack (metrics, traces,
logs)~\cite{b1,b2,b3,b25,b26,b27}, while web-agent benchmarks exercise only
browser-visible state~\cite{b4,b5,b6}. No prior benchmark requires an agent to
reason across both spaces for a single failure. CUJBench addresses Gaps~1 and~2
with two applications chosen for diagnostic contrast: one where failures propagate
through backend service boundaries, and one where failures surface at the browser
boundary. This ensures $\mathcal{E}_{f}$-dominant, $\mathcal{E}_{b}$-dominant, and
cross-modal scenarios are all represented without biasing the corpus toward one
evidence family.

\textbf{Relevance-based scenario difficulty} (Gap~3).
Multiple independent studies show that LLM-based diagnostic agents systematically
under-exploit available evidence: Kim et al.~\cite{b8} find limited telemetry
coverage in 26.9\% of runs and incomplete exploration in 63.9\%, while Riddell et
al.~\cite{b12} and Roy et al.~\cite{b10} independently characterize agents as
fixating on narrow signal even when broader evidence is accessible. If difficulty
is encoded by varying which tools are available, these exploration failures become
indistinguishable from benchmark design: an agent that retrieves only one modality
looks identical whether the other tools were absent or simply ignored. CUJBench
holds the tool interface constant across all scenarios; what varies is which tools
carry decisive signal versus benign background for a given failure. Encoding
difficulty in evidence relevance rather than tool availability makes agent
exploration choices observable: since every tool is always available, a failure to cross modalities is unambiguously an agent behavior and the process-level metrics defined in Section~\ref{sec:eval} can attribute it accordingly.

% ---------------------------------------------------------------
\subsection{Benchmark Harness}
% ---------------------------------------------------------------

CUJBench is designed as a controlled benchmark. The central goal is to make every evaluated failure reproducible, auditable, and stable across agent runs. To this end, the harness separates \textit{scenario generation} from \textit{agent evaluation}: failures are first produced in a fully instrumented environment under controlled conditions, then packaged into fixed scenario snapshots that can be replayed deterministically during benchmarking.

\subsubsection{Test Environment}

The harness runs on two complementary containerized benchmark applications.
OpenTelemetry Demo is a polyglot microservices e-commerce stack instrumented with
distributed tracing, Prometheus metrics, and centralized logging, a
backend-observable environment where failures propagate across service boundaries
and leave evidence in Jaeger, Prometheus, and OpenSearch. Tractor Store is a
micro-frontend e-commerce application~\cite{b50} where failures surface in rendered UI,
browser event flows, and composition boundaries.

CUJ tests are executed via Playwright against both applications. These tests serve
two purposes simultaneously: they encode the user journey as a runnable script and generate the browser-side evidence captured into each scenario package.

\subsubsection{Scenario Schema}

Each scenario is first declared in a registry entry specifying what to run, what
to inject, and what outcome to expect. The core fields are the injection
configuration and an outcome contract: a structured assertion specifying the anticipated user-visible symptom, and the evidence families that must be present in the captured package. Additional fields record modality and difficulty metadata used during evaluation (Figure~\ref{fig:cujbench-e2e}).

This schema is the control plane of the benchmark. It fixes the intended failure
condition before execution and ensures that scenario generation is a controlled
instantiation of a predefined case, not an ad hoc recording process.

\subsubsection{Fault Injection}

The harness supports two injection mechanisms. The generic mechanism is an HTTP layer fault proxy that intercepts traffic and
applies parameterized faults to selected paths, such as status-code overrides or response-body mutations. Because these faults are expressed at the transport
layer, their effects appear naturally in HAR entries, screenshots, and console
output.

Application-specific mechanisms cover fault classes that cannot be expressed at
the HTTP layer. For OpenTelemetry Demo, feature flags activate backend failure
modes that surface primarily in traces, metrics, and logs. For Tractor Store, source-level mutations introduce failure classes such as event-contract drift and cookie-state inconsistencies.

Together, these mechanisms let the benchmark control not only whether a failure
occurs, but also where its diagnostic signal is expected to appear.

\subsubsection{Scenario Snapshot}

After fault injection and CUJ execution, the harness packages all captured
evidence into a self-contained scenario snapshot. The snapshot consolidates
browser-side artifacts with backend observability data and operational context into a stable layout, and pre-computes deterministic per-tool response files from the raw captures.

This packaging step is what makes the benchmark reproducible. The live execution
happens once during scenario generation; evaluation then runs only against the
packaged snapshot, not against a live system. All agents assigned to the same
scenario inspect the same alert, the same artifacts, and the same precomputed
evidence. The snapshot is the atomic unit of the benchmark corpus.

\subsubsection{Agent Tool Interface}

Agents interact with scenario evidence through a tool server that returns
deterministic, cached responses for a given scenario. The interface is fixed
across the benchmark: all full-agent runs receive the same 12 evidence-access
tools plus a submission tool, organized into families that mirror the
$\mathcal{E}_{f}$, $\mathcal{E}_{b}$, $\mathcal{E}_{c}$ partition from
Section~\ref{sec:problem_def}. What varies across scenarios is not the interface
but the diagnostic role of each family: a tool may return decisive signal in one
scenario and a graceful stub in another, making evidence relevance the mechanism
through which difficulty is encoded. Table~\ref{tab:harness-tools} shows how
this varies across the two benchmark applications.

\begin{table*}[t]
\caption{Benchmark tool families. The interface is identical across all scenarios;
what varies is the diagnostic role of each family---decisive signal, benign
background, or intentional stub---depending on the application and scenario.}
\label{tab:harness-tools}
\centering
\small
\setlength{\tabcolsep}{6pt}
\begin{tabular}{@{}p{3.0cm}p{6.5cm}p{6.5cm}@{}}
\toprule
\textbf{Family} & \textbf{Tools} & \textbf{Diagnostic Role by Application} \\
\midrule
Browser-visible &
  \texttt{view\_screenshot} &
  Decisive in both apps; directly exposes rendered failure state. \\[3pt]
Browser/frontend &
  \texttt{get\_cuj\_report}, \texttt{get\_browser\_console},
  \texttt{get\_network\_requests}, \texttt{get\_dom\_snapshot},
  \texttt{get\_browser\_storage}, \texttt{get\_frontend\_spans} &
  Primary signal in Tractor Store; storage and frontend spans are
  typically non-informative stubs for OTel Demo scenarios. \\[3pt]
Backend/context &
  \texttt{search\_logs}, \texttt{get\_traces}, \texttt{get\_error\_rate},
  \texttt{get\_recent\_changes}, \texttt{get\_service\_topology} &
  Primary signal in OTel Demo; error rate and recent changes return
  graceful stubs for Tractor Store. \\[3pt]
Submission &
  \texttt{submit\_root\_cause} &
  Records the final structured diagnosis with cited evidence identifiers. \\
\bottomrule
\end{tabular}
\end{table*}

\subsection{Scenario Generation}

Using the benchmark harness described above, CUJBench constructs its scenario corpus through a staged generation process: define target failure classes, instantiate candidate scenarios, capture their evidence, and curate the final set through multi-agent review.

\subsubsection{LLM-Based Fault Scenario Generation}

The taxonomy in Table~\ref{tab:scenario-taxonomy} defines the coverage target:
every admitted scenario must instantiate a specific family, modality regime, and
difficulty level. Manually authoring candidates to systematically fill this space
is labor-intensive and tends to over-represent familiar fault patterns while
leaving coverage gaps at the boundaries of each family. To address this, we
employ an LLM-based generator agent that enumerates candidates directly from the
taxonomy cells. We instantiate this agent with GPT-5.4~\cite{b49}, leveraging its
code synthesis capability for structured scenario generation at scale. The generator receives the scenario schema, the full
taxonomy with per-family coverage targets, application architecture, CUJ
definitions, and available injection interfaces, and outputs candidate scenarios
in structured JSON form, where each candidate records the target CUJ, intended
fault mechanism, expected failure manifestation, and the taxonomy attributes the
case is meant to exercise. A single generation pass produces 120 candidates,
which then enter the multi-agent generation-evaluation loop described below.

\subsubsection{Multi-Agent Generation-Evaluation Loop}

The multi-agent generation-evaluation loop serves as a quality gate, testing not
only whether a fault can be injected, but whether the resulting packaged evidence
supports a meaningful and reviewable root-cause diagnosis. In
each iteration, the harness captures the candidate's failure evidence,
independent agents assess whether that evidence supports a diagnosable failure,
and the scenario is either accepted, revised, or dropped. Revised candidates
re-enter the loop with adjusted parameters; only those that pass review within
five rounds are admitted into the corpus (Figure~\ref{fig:cujbench-e2e}).

The loop is orchestrated by an interactive coding agent acting as coordinator and optimizer,
which dispatches review tasks, collects reports, revises the scenario, and tracks each candidate's readiness for acceptance, revision, or removal. Within each iteration, three
reviewer roles operate on GPT-5.4, differentiated by system prompt and file
access scope: two independent SRE reviewers and a senior reviewer.

Each iteration begins with harness execution: the harness runs the associated
CUJ under the candidate's fault configuration and packages the resulting alert
and multi-modal evidence into a scenario snapshot $s$.

This snapshot is independently assessed by two SRE reviewer agents, each serving
as an evidence auditor. Operating through bash utilities and benchmark-facing
evidence interfaces, and without access to oracle information, each reviewer
determines whether the snapshot contains sufficient signal to support a plausible
RCA and produces a documented evidence chain recoverable from the available
artifacts.

A senior SRE reviewer, acting as arbiter, compares the two independent reports
and issues an adjudication verdict. The coordinator agent then consumes this verdict alongside the provisional label and evidence chain, and decides among three outcomes: pass the candidate, revise it, or abandon it. When revision is required, the coordinator may adjust fault parameters, change the CUJ, modify the tool interface to surface richer context,
or propose an alternative scenario that better realizes the same taxonomy target.
This self-correcting process ensures that only scenarios with verified,
diagnosable evidence are admitted into the benchmark corpus.

\subsubsection{Ground Truth Annotation}

Annotating failure scenarios requires inspecting multi-modal artifacts to infer component, fault layer, and root-cause type for each case. CUJBench reduces this cost by constructing ground
truth across three layers: automated mapping of the injection specification,
evidence-grounded review from the multi-agent loop, and human validation.

The first layer is extracted from the fault injection configuration. Each
candidate receives an initial root-cause label constructed by mapping the
injection specification to its corresponding RCA fields. For backend flag faults,
this mapping resolves the activated flag to a canonical component, layer, and
fault family; for browser-path proxy faults, it resolves the targeted request
path and fault mode to the affected component and expected failure type; and for
compound scenarios, it composes the partial diagnoses from the constituent
injections into a single compound case with explicit contributing factors.

The second layer is produced by the multi-agent loop: the senior reviewer's
adjudication verdict identifies whether the initial label holds, requires
refinement, or conflicts with the observable evidence, and the coordinator
applies the corresponding update to the annotation based on the recovered
evidence chain.

The third layer is human verification: domain SREs sampled 40 of the 87 accepted scenarios and
inspected the packaged evidence directly, as a real on-call engineer would, to
confirm that the annotated root cause is recoverable from the available
artifacts. Labels that fail this inspection are corrected, or the scenario is
removed from the corpus. Of the 40 sampled, 8 had their annotations corrected.

The result is a label set grounded in what SREs can observe in the
packaged artifacts. The corpus admitted through this process contains 87 of the
120 generated candidates, summarized in Table~\ref{tab:scenario-taxonomy},
spanning 59 OTel Demo cases and 28 Tractor Store cases. Each admitted scenario
carries per-scenario evaluation attributes: modality requirement, root-cause
visibility, diagnostic depth, and difficulty label. These attributes serve as
the stratification variables for the evaluation reported in
Section~\ref{sec:eval}.

\section{Evaluation}
\label{sec:eval}

We evaluate the performance of current LLM agents on browser-to-backend failure diagnosis
and identify where they fail in the diagnostic chain.
We run six LLMs across three evidence-access baselines on the full 87
scenarios, then examine aggregate accuracy, agent behavior, and failure modes in turn.

\subsection{Experimental Setup}

We evaluate six LLMs spanning both proprietary and open-weight
families: Claude Sonnet 4.6, GPT-5.4, and Gemini 3.1 Pro on the proprietary side;
Qwen3-VL-235B-A22B-Instruct, GLM-4.6V, and Llama-4-Scout on the open-weight side,
forming a diverse and representative sample of current LLMs. All models have vision capability
as required by the screenshot evidence in agentic baselines described below.
All runs are issued through OpenRouter for a uniform API surface.

To evaluate how each model performs as evidence access varies, we define three
baseline conditions that hold model, prompt, and submission schema constant while
systematically varying which evidence channels the agent can query. B1 is a
retrieval-only baseline in which all textual scenario evidence, including CUJ reports, browser artifacts, and backend telemetry, is
serialized and the top-ranked chunks retrieved by BM25 against the failure alert are injected as context into a single LLM call with no tool
 loop; screenshots are excluded, as BM25 operates over text only. B2 and B3 are agentic baselines in which the model interacts with scenario
evidence through a ReAct-style tool-calling loop~\cite{b21} capped at 15 turns,
sufficient to invoke each evidence family at least once across the 12-tool
evidence interface. B2 restricts access to the seven browser-visible tools, while B3
provides the full 12-tool surface including backend observability and operational
context. Comparing B2 and B3 under these controlled conditions is designed to isolate the marginal diagnostic contribution of backend evidence, enabling a
direct assessment of the cross-modal hypothesis that motivates CUJBench's two-regime design (Table~\ref{tab:baselines}).

Each evaluation instance is then a triplet of $(\text{scenario},\, \text{model},\, \text{baseline})$,
with 87 scenarios, 6 models, and 3 baselines, the evaluation yields 1{,}566 runs in total.

\begin{table}[htbp]
\caption{Baseline conditions and evidence access. \textit{Visual}: screenshots.
\textit{Browser}: network traces, console, DOM, and frontend spans.
\textit{Backend}: service logs, traces, error rates, topology, and recent changes.}
\label{tab:baselines}
\centering
\small
\setlength{\tabcolsep}{4pt}
\begin{tabular}{@{}lcccc@{}}
\toprule
\textbf{Baseline} & \textbf{Tool Loop} & \textbf{Visual} & \textbf{Browser} & \textbf{Backend} \\
\midrule
B1: Retrieval$^*$ & --         & --         & \checkmark & \checkmark \\
B2: Browser-only  & \checkmark & \checkmark & \checkmark & --         \\
B3: Full agent    & \checkmark & \checkmark & \checkmark & \checkmark \\
\midrule
\multicolumn{5}{l}{\footnotesize $^*$Evidence text-serialized and fed directly as model context; no tool loop.} \\
\bottomrule
\end{tabular}
\end{table}

\subsection{Evaluation Metrics}

CUJBench reports both outcome and process metrics. Outcome accuracy alone can
mask systematic reasoning failures---an agent may produce the correct answer
through incomplete evidence exploration---a pattern documented in prior RCA
agent studies~\cite{b8}.

\textbf{Outcome metrics} (computed over submitted runs only).

\begin{itemize}

\item \textbf{A@1.} Primary accuracy metric. Layer is excluded because it is
typically inferrable from the component and does not independently determine
the remediation action. A@1 is the single-prediction analogue of the
ranking-based A@k used in prior RCA benchmarks~\cite{b3}.
\begin{equation}
\text{A@1} = \mathbb{1}[\hat{c} = c^{*} \wedge \hat{t} = t^{*}]
\label{eq:a1}
\end{equation}

\item \textbf{Component Match (CM), Layer Match (LM), Type Match (TM).}
Per-field exact match after canonicalization (lowercasing and whitespace
normalization).
\begin{equation}
\mathbb{1}[\hat{c}=c^{*}] \qquad \mathbb{1}[\hat{\ell}=\ell^{*}] \qquad \mathbb{1}[\hat{t}=t^{*}]
\label{eq:field_matches}
\end{equation}

\item \textbf{Partial Credit Equal (PCE).} Equal-weight aggregate of the
three field matches.
\begin{equation}
\mathrm{PCE} = \tfrac{1}{3}\bigl(\mathbb{1}[\hat{c}=c^{*}] + \mathbb{1}[\hat{\ell}=\ell^{*}] + \mathbb{1}[\hat{t}=t^{*}]\bigr)
\label{eq:pce}
\end{equation}

\item \textbf{Partial Credit Weighted (PCW).} Component-heavy aggregate,
reflecting that component localization is the primary actionable output,
type narrows the class of remediation needed, and layer carries the least
independent information.
\begin{equation}
\mathrm{PCW} = 0.5\,\mathbb{1}[\hat{c}=c^{*}] + 0.2\,\mathbb{1}[\hat{\ell}=\ell^{*}] + 0.3\,\mathbb{1}[\hat{t}=t^{*}]
\label{eq:pcw}
\end{equation}

\item \textbf{Evidence Recall (ER).} Fraction of ground-truth artifact
identifiers the agent cited, a retrieval-style signal independent of outcome
correctness.
\begin{equation}
  \mathrm{ER} = \begin{cases}
    |\hat{I} \cap I^{*}| \;/\; |I^{*}| & \text{if } I^{*} \neq \emptyset \text{ and run submitted} \\
    \text{N/A} & \text{otherwise}
  \end{cases}
\label{eq:er}
\end{equation}

\end{itemize}

\textbf{Process metrics} (computed over all runs).

\begin{itemize}

\item \textbf{Submission Rate (SR).} Fraction of runs in which the agent
called \texttt{submit\_root\_cause} within budget. A run completion
prerequisite: a low SR reflects failure to conclude rather than an incorrect
answer.

\item \textbf{Tool Coverage (TC).} Recall of mandatory tools from $\tau^{*}$
over the agent's tool-call sequence $\hat{\tau}$, measuring whether the agent
explored the evidence channels required for a well-grounded diagnosis.
\begin{equation}
  \mathrm{TC} = |M \cap \hat{\tau}|\;/\;|M|
\label{eq:tc}
\end{equation}

\item \textbf{Calls.} Total tool invocations per run. A model may issue multiple tool calls per
turn, so Calls can exceed the 15-turn budget; high Calls with low A@1
indicates unfocused exploration.

\item \textbf{Extra Calls (EC).} Invocations outside the allowed set, an
inverse-precision signal for off-path behavior. Let $\mathcal{R} = M \cup A
\cup \{\texttt{submit\_root\_cause}\}$; then:
\begin{equation}
  \mathrm{EC} = |\{\,\sigma \in \hat{\tau} : \sigma \notin \mathcal{R}\,\}|
\label{eq:ec}
\end{equation}

\end{itemize}

TC and EC are consistent with the trajectory-alignment approach of
Cloud-OpsBench~\cite{b3} and address the incomplete-exploration pitfall
identified by Kim et al.~\cite{b8}, extended here to multi-modal evidence.

\textbf{Uncertainty quantification.} We pair A@1 with 95\% confidence intervals
derived from cluster-robust standard errors, clustering at the scenario level.
This accounts for the correlation among runs that share a scenario's intrinsic difficulty
and sizes the intervals accordingly. Per-model intervals are
computed over each model's 261 runs (257 for GLM-4.6V) and reported in Table~\ref{tab:main-results}.

\subsection{Overall Performance}

Table~\ref{tab:main-results} reports all outcome and process metrics across the
six models and three baselines evaluated on the 87-scenario corpus
(1{,}562 completed runs out of 1{,}566 expected; four GLM-4.6V agent-full cells
timed out and are excluded).

\begin{table*}[t]
\caption{Main evaluation results. Best value per column is \textbf{bold}.
Pooled A@1 is each model's A@1 over all 261 runs (257 for GLM-4.6V).}
\label{tab:main-results}
\centering
\setlength{\tabcolsep}{4pt}
\footnotesize
\begin{tabular}{llcrrrrrrrrrrr}
\toprule
\textbf{Model} & \textbf{Baseline} & \textbf{Pooled A@1\,[95\% CI]} &
  \textbf{A@1} & \textbf{SR} &
  \textbf{CM} & \textbf{LM} & \textbf{TM} &
  \textbf{PCE} & \textbf{PCW} &
  \textbf{TC} & \textbf{ER} &
  \textbf{Calls} & \textbf{EC} \\
\midrule
\multirow{3}{*}{Claude Sonnet 4.6}
  & Browser   & \multirow{3}{*}{\textbf{.391}\,[.320,\,.467]} & \textbf{.525} & .975 & .658 & .789 & .789 & \textbf{.746} & \textbf{.724} & .778 & .593 & 10.0 & 3.9 \\
  & Full      &  & .450          & .850 & \textbf{.667} & \textbf{.818} & .606 & .697 & .679 & .893 & \textbf{.731} & 14.8 & 4.7 \\
  & Retrieval &  & .200          & \textbf{1.000} & .436 & .538 & .308 & .427 & .418 & ---  & ---  & ---  & ---  \\
\midrule
\multirow{3}{*}{GPT-5.4}
  & Browser   & \multirow{3}{*}{.211\,[.153,\,.283]} & .310 & \textbf{1.000} & .535 & .616 & .570 & .574 & .562 & .752 & .514 & 10.7 & 4.3 \\
  & Full      &  & .230 & \textbf{1.000} & .360 & .640 & .465 & .488 & .448 & \textbf{.938} & .514 & 13.6 & 4.3 \\
  & Retrieval &  & .092 & \textbf{1.000} & .326 & .430 & .198 & .318 & .308 & ---  & ---  & ---  & ---  \\
\midrule
\multirow{3}{*}{Gemini 3.1 Pro}
  & Browser   & \multirow{3}{*}{.238\,[.188,\,.295]} & .500 & .925 & \textbf{.667} & .722 & \textbf{.806} & .731 & .719 & .704 & .547 & 17.4 & 6.0 \\
  & Full      &  & .100 & .350 & .429 & .786 & .714 & .643 & .586 & .898 & .469 & \textbf{69.7} & \textbf{45.1} \\
  & Retrieval &  & .100 & \textbf{1.000} & .333 & .615 & .256 & .402 & .367 & ---  & ---  & ---  & ---  \\
\midrule
\multirow{3}{*}{GLM-4.6V}
  & Browser   & \multirow{3}{*}{.206\,[.155,\,.268]} & .310 & .828 & .597 & .708 & .514 & .606 & .594 & .655 & .369 & 6.6  & 2.4 \\
  & Full      &  & .229 & .783 & .500 & .766 & .484 & .583 & .548 & .690 & .330 & 6.3  & 2.0 \\
  & Retrieval &  & .080 & .874 & .453 & .573 & .240 & .422 & .413 & ---  & ---  & ---  & ---  \\
\midrule
\multirow{3}{*}{Llama-4-Scout}
  & Browser   & \multirow{3}{*}{.061\,[.039,\,.095]} & .046 & .276 & .261 & .609 & .435 & .435 & .383 & .327 & .163 & 2.4  & 1.1 \\
  & Full      &  & .046 & .207 & .353 & .647 & .294 & .431 & .394 & .297 & .088 & 2.3  & 1.2 \\
  & Retrieval &  & .092 & \textbf{1.000} & .442 & .709 & .221 & .457 & .429 & ---  & ---  & ---  & ---  \\
\midrule
\multirow{3}{*}{Qwen3-VL-235B}
  & Browser   & \multirow{3}{*}{.188\,[.136,\,.254]} & .276 & .770 & .545 & .697 & .485 & .576 & .558 & .656 & .450 & 5.9  & 2.2 \\
  & Full      &  & .184 & .598 & .462 & .788 & .365 & .538 & .498 & .809 & .365 & 8.4  & 2.7 \\
  & Retrieval &  & .103 & \textbf{1.000} & .442 & .744 & .256 & .481 & .447 & ---  & ---  & ---  & ---  \\
\bottomrule
\end{tabular}
\end{table*}

\textbf{Overall difficulty.}
Across 1{,}562 completed runs, the benchmark yields an overall A@1 of 21.6\%.
The best single cell is 52.5\%, reached by Claude Sonnet~4.6 under the
browser-only baseline, with no model approaching ceiling performance.
Thirty of the 87 scenarios produce zero correct diagnoses across every
model--baseline combination that ran them, while the easiest scenario reaches
66.7\%---a wide difficulty range that confirms the corpus spans a meaningful
challenge spectrum.

\textbf{Model ranking.}
Claude Sonnet~4.6 leads by a wide margin with 102 correct diagnoses across 261
runs (39.1\%), followed by a tightly clustered middle band: Gemini~3.1~Pro
(62/261, 23.8\%), GPT-5.4 (55/261, 21.1\%), GLM-4.6V (53/257, 20.6\%), and
Qwen3-VL-235B (49/261, 18.8\%), with Llama-4-Scout trailing far behind
(16/261, 6.1\%).
The scenario-clustered confidence intervals (Table~\ref{tab:main-results})
resolve the ranking into three statistically distinct tiers: Sonnet's interval
sits disjoint above the field, Llama's disjoint below, and the remaining four
form an overlapping band. Within the band, the frontier GPT-5.4 is statistically indistinguishable
from the open-weight GLM-4.6V and Qwen3-VL-235B, indicating that proprietary status alone
confers no consistent advantage on this task.
PCE and PCW partially offset the low A@1 of Qwen3-VL-235B and Llama-4-Scout:
both reach higher per-field accuracy than their A@1 suggests when they do submit
(PCW up to 0.558 and 0.429 respectively), indicating their low A@1 is driven
more by submission failures than by reasoning errors on submitted
runs.

\subsection{Behavioral Analysis}

\textbf{Effect of baseline configuration.}
Aggregating across all six models, agent-browser-only achieves the highest
A@1 (33.0\%), followed by agent-full (20.7\%), with the zero-tool retrieval
baseline (11.1\%) falling substantially below both agentic configurations.
The retrieval collapse is concentrated in Tractor Store scenarios: failures
involving JSON schema drift, MIME-type mismatches, and CSP violations produce
browser-side symptoms whose decisive signal resides in rendered DOM state or browser console output rather than in any
structured text field, making textual retrieval structurally insufficient
for these failure classes.
Despite retrieval's near-perfect submission rate (97.9\%), it lacks the
tool access to surface discriminating evidence; agent-full SR collapses to
62.9\%---models with the full toolset spend their turn limit exploring
without converging on an answer---yet the A@1 ordering is preserved.
Gemini~3.1~Pro illustrates the exploration extreme: its SR drops from 92.5\%
under browser-only to 35.0\% under agent-full, reducing A@1 from 50.0\% to 10.0\%,
the sharpest single-model degradation observed.
This finding challenges the intuition that broader tool access uniformly helps:
browser-only matches or exceeds agent-full A@1 for every model evaluated, and the
effect is starkest for frontier models, where richer evidence access induces more
exploration at the cost of completion.

\textbf{Process metrics as diagnostic signals.}
TC and ER capture orthogonal aspects of agent behavior that A@1 alone
cannot distinguish.
GPT-5.4 under agent-full achieves TC\,=\,0.94, the highest mandatory-tool
coverage, yet its A@1 (23\%) trails Sonnet (45\%), indicating that calling
the right tools is necessary but not sufficient for correct synthesis.
Claude Sonnet~4.6 under agent-full reaches TC\,=\,0.89 while maintaining the
higher A@1, suggesting more efficient tool utilization.
Across submitted runs, component identification (CM) is consistently the
hardest subfield for frontier models (0.33--0.67), while fault-type
identification (TM) is often the easiest under browser-only conditions
(Gemini Browser TM\,=\,0.81; Claude Browser TM\,=\,0.79) but degrades
under agent-full as more ambiguous submissions accumulate.
For smaller models, layer identification (LM) consistently exceeds TM
(Llama LM\,=\,0.61--0.71 vs.\ TM\,=\,0.22--0.44; GLM LM\,=\,0.57--0.77 vs.\ TM\,=\,0.24--0.51).
Claude Sonnet~4.6 achieves the highest ER (0.731 under agent-full) and the
best ER-to-A@1 conversion under agent-full, suggesting superior synthesis
capability given equivalent evidence access.
PCE and PCW add resolution: Claude Sonnet~4.6 Browser leads both aggregates
(PCE\,=\,0.746, PCW\,=\,0.724), narrowly ahead of Gemini~3.1~Pro Browser
(PCE\,=\,0.731, PCW\,=\,0.719), consistent with their close standing on
browser-only A@1 (52.5\% vs.\ 50.0\%).

\textbf{Scenario difficulty.}
Correct diagnoses are unevenly distributed across scenarios.
The easiest scenario tops out at 66.7\% A@1, a single dominant
browser-visible signal (an HTTP 503 surfaced in the HAR), while 30 of the 87
scenarios receive zero correct diagnoses from every model and baseline that
ran them.
Easy scenarios share a common characteristic: a single dominant signal
observable in a standard tool call (e.g., an HTTP 503 in the HAR, or an
anomalous field in a JSON API response).
Hard scenarios require correlating evidence across modalities---browser
console errors, network request timing, and backend spans simultaneously---or
involve fault manifestations (MIME-type mismatches,
JavaScript event naming errors) that are rarely encountered in standard RCA
training corpora.

\textbf{Efficiency vs.\ accuracy.}
Gemini~3.1~Pro under browser-only averages 17.4 tool calls and 6.0 extra calls
per run and consumes roughly 1.2M input tokens on average, yet reaches only
50.0\% A@1, while Sonnet attains a higher 52.5\% at roughly half the token cost
(652K tokens, 10.0 tool calls), indicating that more evidence access and more exploration do
not translate into better diagnoses.
At the other extreme, Llama-4-Scout and Qwen3-VL-235B average only 2--8
tool calls before stalling without a submission, a failure pattern driven by low
SR rather than incorrect synthesis.
Together these results indicate that both over-exploration and premature
abandonment are failure modes distinct from incorrect reasoning, and that
tool-call count and token cost are poor proxies for diagnostic quality on
CUJBench.

\subsection{Agent Failure Modes}

Trajectory-level analysis of the 299 agent non-submissions and the submitted-but-incorrect fraction reveals three structurally distinct failure modes, each arising at a different stage of the diagnostic pipeline.

\textbf{Tool-call formatting instability.}
The dominant failure mode (200 of 299 agent non-submissions) is an inability to emit syntactically valid tool invocations. Llama-4-Scout (132 of 132 non-submissions) and Qwen3-VL-235B (39 of 55 non-submissions) produce coherent natural-language reasoning that correctly identifies investigative next steps but cannot reliably encode those steps as structured tool calls.
Trajectory inspection across affected runs reveals a consistent pattern: after one or a small number of valid tool invocations, the model reverts to producing free-form natural-language reasoning rather than a structured tool call, breaking the output contract the harness requires.
Critically, the plain-text content generated during these failed steps demonstrates coherent diagnostic reasoning: the model continues to identify relevant hypotheses and propose investigative next steps, indicating that the failure is not one of analytical capacity but of output format compliance.
That GPT-5.4 produces zero agent non-submissions on the same harness confirms that formatting instability is a model-specific limitation rather than an inherent property of the diagnostic task.

\textbf{Runaway exploration and context exhaustion.}
In the second failure family, frontier models issue valid tool calls throughout but never converge: evidence accumulates without narrowing query scope until hitting the turn limit or provider context limits.
Gemini~3.1~Pro accounts for 64 of the 99 non-submissions in this family, with failed runs averaging 83.8 tool calls and 4.0M input tokens; two sub-patterns emerge: repeated broad queries that recycle the same tool without filtering, and single-turn fan-outs in which one Gemini run issued 633 tool calls before the provider rejected the subsequent turn for context size.
Claude Sonnet~4.6 shows the same failure in milder form (15 non-submissions, 19.6 tool calls, 3.5M tokens), explaining the agent-full SR collapse: the expanded tool inventory creates more surface area for unfocused queries, and models that converge under browser-only instead exhaust the turn limit exploring with backend tools.

\textbf{Synthesis failures in submitted runs.}
A third failure mode operates silently within the submitted fraction: the agent completes investigation and produces a structured answer, but reasons to the wrong component.
Component identification (CM) is the subfield most resistant to correct prediction across all models (0.33--0.67 for frontier models), while evidence recall (ER) reaches 0.55--0.73 for the best-performing configurations.
The gap between high ER and imperfect A@1 is diagnostically informative: Claude Sonnet~4.6 achieves ER\,=\,0.731 under agent-full yet A@1\,=\,0.450, indicating that the decisive evidence is retrieved in many runs where the final component attribution nonetheless fails.
A complementary signal appears in GPT-5.4, which achieves the highest mandatory-tool coverage (TC\,=\,0.94) but trails Sonnet on A@1 (23\% vs.\ 45\%): calling the right tools is necessary but not sufficient.
Together these patterns suggest that the primary bottleneck for correctly-submitting frontier agents on CUJBench is cross-modal synthesis---mapping retrieved evidence onto a precise component attribution---rather than evidence discovery.

\subsection{Threats to Validity}

\textbf{External validity.} CUJBench is grounded in two open-source full-stack e-commerce applications; absolute performance numbers should be interpreted within this scope. Most faults also originate at the browser or HTTP layer; whether the findings extend to other domains, or to faults rooted in backend internals, is untested. The diagnostic reasoning patterns and failure modes exposed are nonetheless expected to transfer across stacks, as the cross-modal evidence structure and agent failure taxonomy reflect properties of the task, not idiosyncrasies of any single application.

\textbf{Statistical conclusion validity.} Per-model A@1 is reported with 95\% confidence intervals from scenario-clustered standard errors. Only Sonnet and Llama separate from the field at the 95\% level; the four central models' intervals overlap, leaving their relative ordering indicative and not statistically distinguished. Four GLM-4.6V agent-full cells timed out and are excluded from the 1{,}566 expected runs; their number is small enough not to materially affect the reported aggregates.

\textbf{Construct validity.} The primary metric A@1 requires exact match on both component and type and does not credit partial attribution; per-field metrics (CM, LM, TM) provide finer-grained signal but are secondary. Reference trajectories reflect a single expert investigation path among potentially several valid ones, and evaluation prompts use generic templates---model-specific optimization could yield different performance ceilings.

\section{Related Work}\label{sec:related}

\subsection{Agentic RCA Benchmarks}

AIOps benchmarks evaluate AI agents on cloud reliability tasks through two broad
paradigms. Live-environment frameworks such as AIOpsLab~\cite{b1} and
ITBench~\cite{b2} deploy instrumented microservice stacks, inject faults at
runtime, and assess agents across the full incident lifecycle.
Cloud-OpsBench~\cite{b3} introduced the state-snapshot paradigm---capturing
failures once under controlled conditions and replaying them via deterministic
mocked interfaces---eliminating live-cluster dependencies and enabling
process-level evaluation. RCA-focused benchmarks such as OpenRCA~\cite{b24} and
RCAEval~\cite{b25} further target fault localization from backend telemetry;
recent multi-modal extensions combine metrics, logs, and traces within the backend
stack~\cite{b38,b41}, yet Zhang et al.~\cite{b39} find that additional backend sources
do not consistently improve diagnosis outcomes. None of these benchmarks
incorporates browser-visible evidence as diagnostic input. LLM-based incident assistants~\cite{b22} and operational RCA systems~\cite{b23}
address adjacent diagnostic tasks from incident tickets and backend telemetry, but
are not designed as evaluation benchmarks. CUJBench adopts
the snapshot paradigm of Cloud-OpsBench and extends it across the combined browser
and backend evidence surface.

\subsection{Web and GUI Agent Benchmarks}

A parallel thread evaluates LLM agents on visual web and desktop tasks.
WebArena~\cite{b5}, VisualWebArena~\cite{b4}, and OSWorld~\cite{b6} benchmark
agents on task completion over working applications using browser screenshots, DOM
trees, and accessibility structures. Robustness-oriented extensions such as GUI-Robust~\cite{b19} introduce abnormal
interface states, and long-context suites such as VideoWebArena~\cite{b20} target
multi-step execution, but both retain the completion framing: the evaluation
objective is prospective task execution, not retrospective failure attribution.
Failure diagnosis inverts this problem---the application has already broken and the
agent must reason backward from symptoms to root cause---and no web-agent benchmark
instantiates this inversion, nor incorporates backend observability for cross-stack
diagnosis. CUJBench occupies the intersection these two lines
of work leave uncovered: browser-visible evidence combined with backend
observability, in a diagnostic rather than completion framing.

\subsection{Scenario Generation and Agent Failure Analysis}

Prior benchmarks construct scenarios through manual authoring~\cite{b24,b25} or
live fault injection~\cite{b1,b2}; pre-release checks, where present, verify that
faults manifest as specified but not that the captured evidence supports an
independently recoverable diagnosis. CUJBench addresses this through an LLM-based
generator paired with a multi-agent evaluation loop in which candidates are admitted
only after independent reviewer agents verify that packaged evidence supports a
diagnosable root cause.

On the agent side, Kim et al.~\cite{b8} classify 12 pitfall types across 1,675
OpenRCA agent runs, finding that hallucinated data interpretation (71.2\%) and
incomplete exploration (63.9\%) persist across all model tiers, implicating shared
agent architecture rather than individual model limitations. Riddell et al.~\cite{b12}
and Roy et al.~\cite{b10} independently characterize similar fixation failures under
constrained telemetry. Adjacent work on trace debugging (TRAIL~\cite{b13}) and tool-augmented RCA systems
(RCAgent~\cite{b36}) targets related problems but remains confined to backend
evidence and is not designed for cross-modal evaluation. These studies motivate CUJBench's process
metrics---tool coverage, evidence recall, and extra calls---and its evaluation
extends this failure taxonomy to the multi-modal setting, where browser and backend
signals must both be identified and synthesized.

\section{Conclusion}

CUJBench introduces the first benchmark to couple browser-visible failure evidence
with backend observability in a diagnostic task framing, filling a structural blind
spot shared by every existing AIOps and web-agent benchmark. Across 1{,}562 runs spanning
six LLMs, the benchmark yields an overall A@1 of 21.6\% with a ceiling
of 52.5\%, and its trajectory analysis identifies cross-modal
synthesis as the primary bottleneck for correctly-submitting models. The uniformity of this failure across the models we evaluate---spanning capability tiers and providers---suggests
that cross-modal synthesis is a structural limitation of current diagnostic
agents on this task, one that model scale and richer tool APIs alone cannot resolve
without corresponding advances in multi-step hypothesis revision, cross-modal
observation binding, and the evidence-correlation protocols that remain open
problems in agentic systems research.

\end{document}